\documentclass[a4paper,11pt]{article}
\usepackage[dvips]{graphicx}
\usepackage{amssymb}
\usepackage{amsmath}

\newcommand{\plb}{Phys. Lett. B. }

\newcommand{\arx}{arXiv:}

\begin{document}

\title{Kaluza-Klein Pistons with non-Commutative Extra Dimensions}
\author{V.K.Oikonomou\thanks{
voiko@physics.auth.gr}\\
Dept. of Theoretical Physics Aristotle University of Thessaloniki,\\
Thessaloniki 541 24 Greece\\
and\\
T.E.I. Serres} \maketitle

\begin{abstract}
We calculate the scalar Casimir energy and Casimir force for a
$R^3\times N$ Kaluza-Klein piston setup in which the extra
dimensional space $N$ contains a non-commutative 2-sphere,
$S_{FZ}$. The cases to be studied are $T^d\times S_{FZ}$ and
$S_{FZ}$ respectively as extra dimensional spaces, with $T^d$ the
$d$ dimensional commutative torus. The validity of the results and
the regularization that the piston setup offers are examined in
both cases. Finally we examine the 1-loop corrected Casimir energy
for one piston chamber, due to the self interacting scalar field
in the non-commutative geometry. The computation is done within
some approximations. We compare this case for the same calculation
done in Minkowski spacetime $M^D$. A discussion on the
stabilization of the extra dimensional space within the piston
setup follows at the end of the article.
\end{abstract}

\bigskip

\section*{Introduction}

Casimir pistons have received much attention in recent years
\cite{KirstenKaluzaKleinPiston,KirstenPistons,ElizaldePistons,calvacanti,Teo,TeoMassive,vasilis,malakas}.
This is due to the very attractive quantitative features that the
piston setup has. Most of these are concentrated on the fact that
although the Casimir energy of a scalar field between parallel
plates can be singular, in the piston setup the Casimir force is
regular. Thus we can have closed form formulas for the Casimir
force.

\noindent The Casimir piston can be realized by three parallel
plates in which the one in the middle can be movable. Suppose
between the plates a scalar field exists. The boundary conditions
on the plates can be either Dirichlet or Neumann. In most cases it
was found that the movable plate is attracted to the nearest end.
This holds for rectangular cavities. However when closed cylinder
with arbitrary cross section is considered with Dirichlet piston
and Neumann walls, a repulsive force was found \cite{hybrid}. On
the contrary a three dimensional piston with all surfaces
perfectly conducting is attracted to the closest wall
\cite{marakefsky}.

\noindent The parallel plates Casimir force taking into account
extra compact spatial dimensions has been used to put restrictions
on the size of the extra dimensions \cite{perivolaropoulos}. Thus
the calculation of Casimir force on a piston in the presence of
compact extra dimensions is a very interesting task. The
incorporation of additional Kaluza-Klein dimensions in pistons was
made in \cite{KirstenKaluzaKleinPiston,Cheng,ElizaldePistons}. In
reference \cite{KirstenKaluzaKleinPiston} the scalar Casimir force
was computed for a three dimensional piston with arbitrary
Kaluza-Klein extra dimensions. It was found that for both Neumann
and Dirichlet boundary conditions, the piston is attracted to the
closest wall. This statement was true for arbitrary cross section
of the piston and also for arbitrary geometry and topology of the
extra compact dimensions. This is a very serious argument. As in
most cases the scalar Casimir energy was singular, but due to the
piston setup the Casimir force was finite and found in closed form
using zeta regularization techniques \cite{elizalde,kirstenbook}.

\noindent The argument that the scalar Casimir force for the
piston is independent on the topology and the geometry of the
compact extra dimensions is very serious as we said. This holds
for compact Riemann spaces with arbitrary curvature. In this
article we shall check this argument in the case where the extra
dimensional manifold consists of non-commuting extra dimensions.

\noindent The study for non-commuting extra dimensions was
motivated by string theories studies of D0 branes in constant
background Ramond-Ramond (RR) Chern-Simons field
\cite{huang,myers,nam}. It was shown that this system leads to a
non commutative 2-sphere which is known as fuzzy sphere. We shall
discuss on this later on.

\noindent Beside the above reasoning for using them, non-commuting
extra dimensions can solve a problem in field theories with
compact extra dimensions. In most cases one must explain why extra
dimensions are so small compared to the other three spatial
dimensions \cite{nam}. Also explain why the extra compact
dimensions did not inflate as the other three dimensions. This was
first dealt by Appelquist and Chodos \cite{chodos}. They found a
negative Casimir energy in the bulk five dimensional model they
studied. This resulted in the shrinking of the extra dimension. In
addition in order the extra dimension does not collapse to the
Planck scale, one must find a mechanism for radius stabilization.
It is very natural to expect that radius stabilization of our
field theory can easily occur if there exists a natural minimum
length scale in our theory. One probable scale is the Planck
scale. This is not so good news for extra dimensions experiments.
However if we assume that the extra dimensions can be of Tev scale
then another intrinsic length must be found. Non-commutative extra
dimensions offer this possibility. Non-commutativity naturally
introduces a minimum volume, that of the Moyal cell, which is
proportional to the non-commutativity parameter (we shall discuss
on these issues in the following sections) which has dimension of
length \cite{nam}. Thus if someone considers the ordinary compact
extra dimensions and additionally some extra dimensional
non-commutative compact space (or just the non-commutative space),
this could lead to the stabilization of the extra spaces. This is
a strong motivation for using non-commutative extra dimensions
\cite{nam}. The stabilization of the extra dimensional space in
the case of the piston is a very interesting task but we shall not
deal with it in this article but we will discuss some aspects at
the end of the article. We shall only examine the question if the
scalar Casimir force for the piston is independent on the topology
and the geometry of the compact extra dimensions, in the case
non-commutative extra dimensions exist.

\noindent Let us mention here that the non-commutativity spoils
Lorentz invariance. However having non-commutativity in compact
extra spatial dimensions is not a serious opposition to
observations so far \cite{nam}.

\noindent In this paper we shall study the massive self
interacting $\Phi^4$ scalar field theory in a piston setup. We
shall assume that the spacetime has additional compact dimensions
with topology $T^d\times S_{FZ}$, with $T^d$ the commutative $d$
dimensional torus and $S_{FZ}$ the 2 dimensional non-commutative
sphere, or the fuzzy sphere. The total space has topology
$R^3\times T^d\times S_{FZ}$. The piston will be assumed to have
two infinite dimensions and also the two plates will be at
distance $L$. The piston will be $\alpha$ from the one plate and
$L-\alpha$ from the other (we follow reference
\cite{KirstenKaluzaKleinPiston}). The distance from the plate
shall be generally denoted $a$ and apply the general result that
we will find for the $\alpha$ and $L-\alpha$ cases. The scalar
field shall be assumed to have Dirichlet boundary conditions on
the plates. Our purpose is to find the Casimir energy and Casimir
force for the self interacting massive scalar field in the piston
plates setup, at tree and at 1-loop level. The 1-loop level
contributions will be calculated in a specific approximation which
serves to obtain an analytic result.

\noindent The outline is as follows. In section 1 we shall review
the necessary features of the non-commutative 2-sphere. The
presentation will be enough detailed so the presentation is self
contained (we follow references \cite{nam,myers,huang,tsexoi}). In
section 2 the eigenvalues of the self interacting $\Phi^4$ scalar
field in the piston setup shall be given. In section 3 and 4 we
compute the Casimir energy for the piston setup and also the
Casimir force for $T^d\times S_{FZ}$ and $S_{FZ}$ as extra
dimensional space. In section 5 we add one loop corrections to the
Casimir energy. In section 6 the conclusions follow.

\section{The Fuzzy Sphere, Matrix Models and D0 Branes}

\subsection{D0 Branes Action in the Presence of RR Field}
To start we shall present the string theory application of the
fuzzy sphere that strongly motivates non-commutative geometry
\cite{myers}. Consider N D0 branes in a constant background RR
four form field, $F^{(4)}_{tijk}=-2\lambda_{N}\epsilon_{ijk}$. The
Born-Infeld action for the D0 branes (that is for Dirichlet branes
with zero spatial dimensions) in the presence of the constant RR
field (in leading order within the matrix theory approximation) is
\cite{myers,huang},
\begin{equation}\label{actionbrane}
S=T_{0}\mathrm{Tr}\Big{(}\frac{1}{2}\dot{X}_i^2+\frac{1}{4}[X_i,X_j]{\,}[X_i,X_j]-\frac{i}{3}\lambda_N\epsilon_{ijk}X_i{\,}[X_j,X_k]\Big{)}
\end{equation}
where the last two terms are the leading order Born-Infeld
potential. In relation (\ref{actionbrane}) the $X_i$ are $N\times
N$ matrices and $T_0$ is the zero brane tension \cite{myers}. The
static equations of motion for the above action are,
\begin{equation}\label{equationofmotion}
\Big{[}X_j,\Big{(}[X_i,X_j]-i\lambda_N\epsilon_{ijk}X_i[X_j,X_k]\Big{)}\Big{]}.
\end{equation}
The energy of the extrema is,
\begin{equation}\label{energyextrema}
E=-T_0\mathrm{Tr}\Big{(}+\frac{1}{4}[X_i,X_j]{\,}[X_i,X_j]-\frac{i}{3}\lambda_N\epsilon_{ijk}X_i{\,}[X_j,X_k]\Big{)}
\end{equation}
The above equation clearly admits the commuting matrices solution,
\begin{equation}\label{commuting}
[X_i,X_j]=0
\end{equation}
with energy $E=0$. A non commuting ansatz solution for equation
(\ref{equationofmotion}) is given by the $N\times N$ matrix
representation of $SU(2)$ algebra,
\begin{equation}\label{noncom}
[X_i,X_j]=i\lambda_N\epsilon_{ijk}X_k
\end{equation}
with,
\begin{equation}\label{jma}
X_i=\lambda_NJ_i,
\end{equation}
The matrices $J_i$ define the $N$ dimensional $SU(2)$
representation we mentioned previously, and are labelled by the
total spin $N/2$. The non-commutativity parameter is of dimension
length and can be taken positive. Within the D0 interaction
context, the solution minimizing equation (\ref{equationofmotion})
will be the irreducible $N\times N$ $SU(2)$ representation given
by the non-commuting matrices $X_i$, satisfying equation
(\ref{noncom}). As we will describe shortly, relation
(\ref{noncom}) suggests the fuzzy sphere geometry. Indeed the
matrices satisfy,
\begin{equation}\label{sphere}
X_1^2+X_2^2+X_3^2=r^2_N.
\end{equation}
We shall discuss on the meaning of $r_N$ shortly. Thus we saw that
the appearance of non-commutativity in string theory arrives
naturally within the N D0 brane action in the presence of a
constant RR Chern Simons term. The ground state of this system has
the fuzzy sphere geometry. Indeed the non-commutative setup we
described has minimum energy compared to all other representations
of the $SU(2)$ algebra. So the D0 brane system condense into a
non-commutative configuration. This can be shown to be a bound
state of a spherical D2 brane and $N$ D0 branes (for details see
\cite{myers}).

\subsection{The Fuzzy Sphere}

The infinite dimensional algebra $A_{\infty}$ contains all the
information about the standard sphere $S^2$. It is generated by
$x=(x_1,x_2,x_3)$ $\epsilon$ $R^3$, with,
\begin{equation}\label{aapeiro}
[x_i,x_j]=0,
\end{equation}
and
\begin{equation}\label{alli2}
\sum_{i=1}^3x_i^2=\rho^2
\end{equation}
The non-commutative sphere is connected with the non-commutative
algebra $A_N$ \cite{huang,tsexoi}. This algebra is generated by
$x=(\widehat{x}_1,\widehat{x}_2,\widehat{x}_3)$ with,
\begin{equation}\label{an}
[\widehat{x}_i,\widehat{x}_j]=i\lambda_N\epsilon_{ijk}\widehat{x}_k,
\end{equation}
and as before,
\begin{equation}\label{alli2n}
\sum_{i=1}^3\widehat{x}_i^2=\rho^2
\end{equation}
The parameter $\lambda_N$ characterizes the non-commutativity as
we seen before. Writing the above in terms of
$\widehat{J}_i=\frac{1}{\lambda_N}\widehat{x}_i$,
\begin{equation}\label{cambio}
[\widehat{J}_i,\widehat{J}_j]=i\epsilon_{ijk}\widehat{J}_k.
\end{equation}
Also the Casimir operator $C$ of the group is,
\begin{equation}\label{casimirgroup}
C=\sum_{i=1}^3\widehat{J}_i^2=\rho^2\lambda^{-2}_{N}
\end{equation}
For the case in interest (that is for the $N\times N$ irreducible
representation of the $SU(2)$ algebra), the Casimir operator takes
the value,
\begin{equation}\label{ircasim}
C=\frac{N}{2}\Big{(}\frac{N}{2}+1\Big{)}.
\end{equation}
Thus the $\lambda$ and $N$ are related by,
\begin{equation}\label{aminaca}
\rho
\lambda^{-1}_{N}=\sqrt{\frac{N}{2}\Big{(}\frac{N}{2}+1\Big{)}}.
\end{equation}
We saw earlier the parameter $r_N$. Now for the greater
irreducible representation, this parameter is equal to,
\begin{equation}\label{aminacasideshow}
r_N
\lambda^{-1}_{N}=\sqrt{\frac{N}{2}\Big{(}\frac{N}{2}+1\Big{)}}.
\end{equation}
We can see that the radius of the non-commutative 2-sphere is
finite and depends on the representation of the $SU(2)$ algebra
through the parameter $N$. Also it depends on the
non-commutativity parameter $\lambda_N$. Thus the radius of the
fuzzy sphere is fixed and quantized in terms of the
non-commutativity parameter.

\noindent Finally the Laplacian on the non-commutative sphere is
given by $\Delta=J_1^{{\,}2}+J_2^{{\,}2}+J_3^{{\,}2}$. We shall
use this in the following when we obtain the eigenvalues of the
total space.

\section{Eigenvalues and eigenfunctions}

Consider the scalar field $\Phi$ in space topology $R^3\times
T^d\times S_{FZ}$, where as we seen $S_{FZ}$ is the fuzzy sphere.
In this section we compute the eigenvalues at tree level (that is
Kaluza-Klein spectrum). The scalar field is supposed to obey
Dirichlet boundary conditions on the piston boundaries, that is,
\begin{equation}\label{bc}
\Phi(x^{D-1},0,y)=\Phi(x^{D-1},\alpha,y)=\Phi(x^{D-1},L,y)=0.
\end{equation}
In the above equation, $D$ is the total dimensionality of the
non-compact dimensions (in the end $D=4$), $x$ denotes the
non-compact dimensions and $y$ denotes the $d$ dimensional torus
coordinates. We suppose the torus has equal radii for all the
dimensions. The non-commutative sphere is described by the
non-commutative coordinates $X=(X_1,X_2,X_3)$ of relation
(\ref{noncom}) which are the $N$ dimensional irreducible
representation of the $SU(2)$ algebra. The radius $r_N$ is fixed
as we seen before, see relation (\ref{aminacasideshow}). The
action for the scalar field reads,
\begin{equation}\label{actions}
S=\int \mathrm{d}^Dx\int \mathrm{d}^dy \int_{Tr}
\Big{[}\frac{1}{2}\Phi\big{(}\square_{x}+\square_{y}+\Delta+m^2\big{)}\Phi+\frac{g}{4!}\Phi^4\Big{]}.
\end{equation}
In the above $\Delta=\sum J_i^{{\,}2}$ is the Laplacian, where
$J_i$ as in relation (\ref{cambio}). Also the integral $\int_{Tr}$
is defined as an integration over the fuzzy sphere as follows,
\begin{equation}\label{fuzzyinte}
\int_{Tr}F(X)= \frac{4\pi}{N+1}\mathrm{Tr}[F(X)]
\end{equation}
where $X$ is defined before. The scalar field $\Phi$ has the
following harmonic expansion,
\begin{equation}\label{harmexpans}
\Phi(x,y,X_i)=\int
\frac{\mathrm{d}^{D-1}p}{(2\pi)^{D-1}}\sum_{n=1}^{\infty}\sum_{\vec{n}}\sum_{J,m}b_m^J(\vec{p},\vec{n})e^{i\vec{p}\vec{x}}e^{i2\pi\vec{n}\vec{y}/R}Y_m^J\sin(\frac{n\pi}{a}).
\end{equation}
In the above $\vec{n}=(n_1,n_2,..,n_d)$ and $Y_m^J$ the usual
spherical harmonics. The eigenfrequencies of the scalar field in
the cavity at tree order are given by,
\begin{equation}\label{eigentree}
(w_{\vec{n},\vec{p}}^{J,n})^2=\vec{p}^{{\,}2}+(n\pi/a)^2+(\vec{n}/R)^2+J(J+1)+m^2.
\end{equation}
Also in the case of the fuzzy sphere as the extra dimensional
space the eigenvalues are,
\begin{equation}\label{eigesp}
(w_{\vec{p}}^{J,n})^2=\vec{p}^{{\,}2}+(n\pi/a)^2+J(J+1)+m^2.
\end{equation}

\section{Casimir piston for $T^d\times S_{FZ}$ extra dimensional space}

Let us calculate the Casimir energy and Casimir force for the
piston setup in the case the extra dimensional is the
d-dimensional commutative torus and the fuzzy sphere. We shall do
the calculations for general $s$ and $D$. This will give us the
opportunity to have a general form of the results for all space
dimensions. In the end we have in mind that $s=-\frac{1}{2}$ and
$D=3$. The Casimir energy reads,
\begin{align}\label{vlad}
&E_c(s,a)=
\\& \notag \frac{1}{4\pi^2}\int
\mathrm{d}^{D-1}p\sum_{J=0}^N\sum_{n=1}^{\infty}\sum^{\infty}_{n_1,n_2,...,{\,}n_d=-\infty}\Big{[}\sum_{k=1}^{D-1}p_k^2+(\frac{n\pi}{a})^2+\sum_{k=1}^d(\frac{n_k}{R})^2+J(J+1)+m^2\Big{]}^{-s}.
\end{align}
Upon integrating over the continuous dimensions using,
\begin{equation}\label{feynman}
\int
\mathrm{d}k^{D-1}\frac{1}{(k^2+A)^s}=\pi^{\frac{D-1}{2}}\frac{\Gamma(s-\frac{D-1}{2})}{\Gamma(s)}\frac{1}{A^{s-\frac{D-1}{2}}}
\end{equation}
relation (\ref{vlad}),
\begin{align}\label{pordoulis}
&E_c(s,a)=\frac{1}{4\pi^2}\pi^{\frac{D-1}{2}}\frac{\Gamma(s-\frac{D-1}{2})}{\Gamma(s)}\times
\\& \notag\sum_{J=0}^N\sum_{n=1}^{\infty}\sum^{\infty}_{n_1,n_2,...,{\,}n_d=-\infty}\Big{[}(\frac{n\pi}{a})^2+\sum_{k=1}^d(\frac{n_k}{R})^2+J(J+1)+m^2\Big{]}^{\frac{D-1}{2}-s}.
\end{align}
Using the identity,
\begin{equation}\label{aplodiplo}
\sum_{n=1}^{\infty}f(n)=\frac{1}{2}\Big{(}\sum_{n=-\infty}^{\infty}f(n)-f(0)\Big{)}
\end{equation}
which holds for an even function of $n$, which is our case. Thus
relation (\ref{pordoulis}) can be written,
\begin{align}\label{pordoulis12}
&E_c(s,a)=\frac{1}{8\pi^2}\pi^{\frac{D-1}{2}}\frac{\Gamma(s-\frac{D-1}{2})}{\Gamma(s)}
\times\\& \notag
\Big{(}\sum_{J=0}^N\sum_{n=-\infty}^{\infty}\sum^{\infty}_{n_1,n_2,...,{\,}n_d=-\infty}\Big{[}(\frac{n\pi}{a})^2+\sum_{k=1}^d(\frac{n_k}{R})^2+J(J+1)+m^2\Big{]}^{\frac{D-1}{2}-s}\\&
\notag
-\sum_{J=0}^N\sum^{\infty}_{n_1,n_2,...,{\,}n_d=-\infty}\Big{[}(\sum_{k=1}^d(\frac{n_k}{R})^2+J(J+1)+m^2\Big{]}^{\frac{D-1}{2}-s}\Big{)}
\end{align}
The above relation can be written in terms of the inhomogeneous
Epstein zeta function
\cite{elizalde,kirstenbook,kirsten12,kirsten14},
\begin{equation}\label{cativoepsteinzeta}
Z_d^{v^2}\Big{(}s;w_1,...,w_d\Big{)}=\sum_{n_1...n_N=-\infty}^{\infty}[w_1n_1^2+...+w_dn_d^2+v^2]^{-s}.
\end{equation}
as follows,
\begin{align}\label{pordoulis13}
&E_c(s,a)=\frac{1}{8\pi^2}\pi^{\frac{D-1}{2}}\frac{\Gamma(s-\frac{D-1}{2})}{\Gamma(s)}
\Big{[}\sum_{J=0}^NZ_{d+1}^{w^2_1}(s-\frac{D-1}{2};\frac{\pi^2}{a^2},\frac{1}{R^2},...,\frac{1}{R^2})
\\&\notag-\sum_{J=0}^NZ_{d}^{w^2_1}(s-\frac{D-1}{2};\frac{1}{R^2},...,\frac{1}{R^2})\Big{]}
\end{align}
with $w_1^2=J(J+1)+m^2$. Upon using the following expansion for
the inhomogeneous Epstein zeta
\cite{elizalde,kirstenbook,kirsten12,kirsten14},
\begin{align}\label{kepst1}
&Z_d^{w^2}\Big{(}s;a_1,...,a_d\Big{)}=\frac{\pi^{\frac{d}{2}}}{\sqrt{a_1{\,}a_2...a_d}}\frac{\Gamma(s-\frac{d}{2})}{\Gamma(s)}{\,}w^{d-2s}\\&
\notag+\frac{2\pi^{s}w^{\frac{d}{2}-s}}{\Gamma(s)\sqrt{a_1{\,}a_2...a_d}}\sum^{\infty'}_{n_1,n_2,...,{\,}n_d=-\infty}\Big{[}\sum_{k=1}^d\frac{n_k^2}{a_k}\Big{]}^{\frac{1}{2}(s-\frac{d}{2})}\times
\\& \notag K_{\frac{d}{2}-s}\Big{(}2\pi w\Big{[}\sum_{k=1}^d\frac{n_k^2}{a_k}\Big{]}^{\frac{1}{2}}\Big{)}
\end{align}
the Casimir energy (\ref{pordoulis13}) reads,
\begin{align}\label{pordoulis1334}
&E_c(s,a)=\frac{1}{8\pi^2}\pi^{\frac{D-1}{2}}\frac{\Gamma(s-\frac{D-1}{2})}{\Gamma(s)}
\times \\& \notag
\sum_{J=0}^N\Bigg{\{}\frac{R^d{\,}a{\,}\pi^{d}{\,}\Gamma(s-\frac{D-1}{2}-\frac{d+1}{2})}{\Gamma(s-\frac{D-1}{2})}{\,}\Big{(}\sqrt{J(J+1)+m^2}{\,}\Big{)}^{d+1-2(s-\frac{D-1}{2})}\\&
\notag+\frac{2\pi^{s-\frac{D-1}{2}}{\,}R^d{\,}a{\,}\Big{(}\sqrt{J(J+1)+m^2}{\,}\Big{)}^{\frac{d+1}{2}-(s-\frac{D-1}{2})}}{\Gamma(s-\frac{D-1}{2})}
\times \\& \notag
\sum^{\infty'}_{n,n_1,n_2,...,{\,}n_d=-\infty}\Big{[}\sum_{k=1}^d(n_kR)^2+\Big{(}\frac{n{\,}a}{\pi}\Big{)}^2\Big{]}^{\frac{1}{2}(s-\frac{D-1}{2}-\frac{d+1}{2})}\times
\\& \notag K_{\frac{d+1}{2}-(s-\frac{D-1}{2})}\bigg{(}2\pi \sqrt{J(J+1)+m^2}{\,}\Big{[}\sum_{k=1}^d(n_kR)^2+\Big{(}\frac{n{\,}a}{\pi}\Big{)}^2\Big{]}^{\frac{1}{2}}\bigg{)}\\& \notag
-\frac{R^d{\,}\pi^{d/2}\Gamma(s-\frac{D-1}{2}-\frac{d}{2})}{\Gamma(s-\frac{D-1}{2})}{\,}\Big{(}\sqrt{J(J+1)+m^2}{\,}\Big{)}^{d-2(s-\frac{D-1}{2})}\\&
\notag-\frac{2\pi^{s-\frac{D-1}{2}}R^d\Big{(}\sqrt{J(J+1)+m^2}{\,}\Big{)}^{\frac{d}{2}-(s-\frac{D-1}{2})}}{\Gamma(s-\frac{D-1}{2})}\sum^{\infty'}_{n_1,n_2,...,{\,}n_d=-\infty}\Big{[}\sum_{k=1}^d(n_kR)^2\Big{]}^{\frac{1}{2}(s-\frac{D-1}{2}-\frac{d}{2})}\times
\\& \notag K_{\frac{d}{2}-(s-\frac{D-1}{2})}\bigg{(}2\pi \sqrt{J(J+1)+m^2}{\,}\Big{[}\sum_{k=1}^d(n_kR)^2\Big{]}^{\frac{1}{2}}\bigg{)}\Bigg{\}}
\end{align}
It is obvious that the above expression contains singularities for
the case $D=3$ and $s=-\frac{1}{2}$. The anomalies are contained
to the two gamma functions, namely,
$\Gamma(s-\frac{D-1}{2}-\frac{d+1}{2})$ and
$\Gamma(s-\frac{D-1}{2}-\frac{d}{2})$. The first is singular for
any $d=$even value and the second is singular for any $d$=odd
value. However the piston setup offers a nice way out from these
singularities. Firstly for computing the force we shall need the
derivative over $a$, which will eliminate the last two terms of
(\ref{pordoulis1334}). Thus one singularity can be avoided. Also
the other singularity can be avoided only for the piston setup,
due to linearity to $a$. Taking the derivative over $a$ and adding
the contributions from the two chambers, namely the $a$ and the
$L-a$, the singularity is cancelled. Thus the total Casimir energy
for the piston configuration is regular. This is similar to the
case studied in \cite{kinezoi}. The reason behind this
cancellation is of course the piston configuration and the
independence of the 'non-homogeneity' parameter of the sums (in
our case $J(J+1)+m^2$) from $a$. We shall see in a following
section that when the 'non-homogeneity' parameter is $a$
dependent, the Casimir energy and the Casimir force are singular
even for the piston setup.

\noindent Adding the contributions from the two chambers, the
Casimir force,
\begin{equation}\label{tcf}
F_c=-\frac{\partial E_{c}(s,L-\alpha)}{\partial
\alpha}-\frac{\partial E_{c}(s,\alpha)}{\partial \alpha},
\end{equation}
is equal to,
\begin{align}\label{casmirforce}
&F_c(s)=-\frac{1}{8\pi^2}\pi^{\frac{D-1}{2}}\frac{\Gamma(s-\frac{D-1}{2})}{\Gamma(s)}
\times \\& \notag \sum_{J=0}^N\frac{\partial}{\partial
a}\Bigg{\{}\frac{2\pi^{s-\frac{D-1}{2}}{\,}R^d{\,}a{\,}\Big{(}\sqrt{J(J+1)+m^2}{\,}\Big{)}^{\frac{d+1}{2}-(s-\frac{D-1}{2})}}{\Gamma(s-\frac{D-1}{2})}
\times \\& \notag
\sum^{\infty'}_{n,n_1,n_2,...,{\,}n_d=-\infty}\Big{[}\sum_{k=1}^d(n_kR)^2+\Big{(}\frac{n{\,}a}{\pi}\Big{)}^2\Big{]}^{\frac{1}{2}(s-\frac{D-1}{2}-\frac{d+1}{2})}\times
\\& \notag K_{\frac{d+1}{2}-(s-\frac{D-1}{2})}\bigg{(}2\pi \sqrt{J(J+1)+m^2}{\,}\Big{[}\sum_{k=1}^d(n_kR)^2+\Big{(}\frac{n{\,}a}{\pi}\Big{)}^2\Big{]}^{\frac{1}{2}}\bigg{)}
 \\& \notag +\frac{2\pi^{s-\frac{D-1}{2}}{\,}R^d{\,}(L-a){\,}\Big{(}\sqrt{J(J+1)+m^2}{\,}\Big{)}^{\frac{d+1}{2}-(s-\frac{D-1}{2})}}{\Gamma(s-\frac{D-1}{2})}
\times \\& \notag
\sum^{\infty'}_{n,n_1,n_2,...,{\,}n_d=-\infty}\Big{[}\sum_{k=1}^d(n_kR)^2+\Big{(}\frac{n{\,}(L-a)}{\pi}\Big{)}^2\Big{]}^{\frac{1}{2}(s-\frac{D-1}{2}-\frac{d+1}{2})}\times
\\& \notag K_{\frac{d+1}{2}-(s-\frac{D-1}{2})}\bigg{(}2\pi \sqrt{J(J+1)+m^2}{\,}\Big{[}\sum_{k=1}^d(n_kR)^2+\Big{(}\frac{n{\,}(L-a)}{\pi}\Big{)}^2\Big{]}^{\frac{1}{2}}\bigg{)}\Bigg{\}}
\end{align}
From the beginning we considered the non-commutativity parameter
to be of order 1. This will be our fundamental length scale and we
shall use this to compare length scales. We shall now compute the
above Casimir force in special cases. The most interesting case is
when $a\gg R$ and $L-a\gg R$. In this case $a\gg 1$ and $R\ll 1$.
We shall take the asymptotic limit of the Bessel functions after
taking the derivative of the above expressions.

\noindent Using,
\begin{equation}\label{81}
\frac{K_{\nu }(xz)}{z^{\nu }}=\frac{1}{2}\int_{0}^{\infty }
\frac{e^{-\frac{x}{2}(t+\frac{z^{2}}{t})}}{t^{\nu +1}}dt ,
\end{equation}
the derivative of $\frac{K_{\nu }(xz)}{z^{\nu }}$ appearing in
(\ref{casmirforce}) is equal to,
\begin{equation}\label{derivative}
\frac{\partial}{\partial z}\Big{(}\frac{K_{\nu
}[x(az^2+b)^c]}{\big{[}(az^2+b)^c\big{]}^{\nu
}}\Big{)}=-x{\,}z{\,}c{\,}a(az^2+b)^{c-1}\frac{K_{\nu+1
}[x(az^2+b)^c]}{\big{[}(az^2+b)^c\big{]}^{\nu+1 }}
\end{equation}
Using (\ref{derivative}) the total Casimir force for the two
chambers of the piston reads,
\begin{align}\label{aftercasimir}
&F_c(s)=-\frac{1}{8\pi^2}\pi^{\frac{D-1}{2}}\frac{\Gamma(s-\frac{D-1}{2})}{\Gamma(s)}
\times \\& \notag
\sum_{J=0}^N\Bigg{\{}\frac{2\pi^{s-\frac{D-1}{2}}{\,}R^d{\,}{\,}\Big{(}\sqrt{J(J+1)+m^2}{\,}\Big{)}^{\frac{d+1}{2}-(s-\frac{D-1}{2})}}{\Gamma(s-\frac{D-1}{2})}
\times \\& \notag
\sum^{\infty'}_{n,n_1,n_2,...,{\,}n_d=-\infty}\Big{[}\sum_{k=1}^d(n_kR)^2+\Big{(}\frac{n{\,}a}{\pi}\Big{)}^2\Big{]}^{\frac{1}{2}(s-\frac{D-1}{2}-\frac{d+1}{2})}\times
\\& \notag K_{\frac{d+1}{2}-(s-\frac{D-1}{2})}\bigg{(}2\pi \sqrt{J(J+1)+m^2}{\,}\Big{[}\sum_{k=1}^d(n_kR)^2+\Big{(}\frac{n{\,}a}{\pi}\Big{)}^2\Big{]}^{\frac{1}{2}}\bigg{)}
 \\& \notag -\frac{2\pi^{s-\frac{D-1}{2}}{\,}R^d{\,}\Big{(}\sqrt{J(J+1)+m^2}{\,}\Big{)}^{\frac{d+1}{2}-(s-\frac{D-1}{2})}}{\Gamma(s-\frac{D-1}{2})}
\times \\& \notag
\sum^{\infty'}_{n,n_1,n_2,...,{\,}n_d=-\infty}\Big{[}\sum_{k=1}^d(n_kR)^2+\Big{(}\frac{n{\,}(L-a)}{\pi}\Big{)}^2\Big{]}^{\frac{1}{2}(s-\frac{D-1}{2}-\frac{d+1}{2})}\times
\\& \notag K_{\frac{d+1}{2}-(s-\frac{D-1}{2})}\bigg{(}2\pi \sqrt{J(J+1)+m^2}{\,}\Big{[}\sum_{k=1}^d(n_kR)^2+\Big{(}\frac{n{\,}(L-a)}{\pi}\Big{)}^2\Big{]}^{\frac{1}{2}}\bigg{)}\Bigg{\}}
\\& \notag -\frac{2\pi^{s-\frac{D-1}{2}}{\,}R^d{\,}a^2{\,}\Big{(}\frac{d+1}{2}-(s-\frac{D-1}{2})\Big{)}\Big{(}\sqrt{J(J+1)+m^2}{\,}\Big{)}^{\frac{d+1}{2}+1-(s-\frac{D-1}{2})}}{\Gamma(s-\frac{D-1}{2})}
\times \\& \notag
\sum^{\infty'}_{n,n_1,n_2,...,{\,}n_d=-\infty}\frac{n^2}{\pi^2}\Big{[}\sum_{k=1}^d(n_kR)^2+\Big{(}\frac{n{\,}a}{\pi}\Big{)}^2\Big{]}^{\frac{1}{2}(s-\frac{D-1}{2}-\frac{d+1}{2}-\frac{3}{2})}\times
\\& \notag K_{1+\frac{d+1}{2}-(s-\frac{D-1}{2})}\bigg{(}2\pi \sqrt{J(J+1)+m^2}{\,}\Big{[}\sum_{k=1}^d(n_kR)^2+\Big{(}\frac{n{\,}a}{\pi}\Big{)}^2\Big{]}^{\frac{1}{2}}\bigg{)}
 \\& \notag +\frac{2\pi^{s-\frac{D-1}{2}}{\,}R^d{\,}(L-a)^2{\,}\Big{(}\frac{d+1}{2}-(s-\frac{D-1}{2})\Big{)}\Big{(}\sqrt{J(J+1)+m^2}{\,}\Big{)}^{1+\frac{d+1}{2}-(s-\frac{D-1}{2})}}{\Gamma(s-\frac{D-1}{2})}
\times \\& \notag
\sum^{\infty'}_{n,n_1,n_2,...,{\,}n_d=-\infty}\frac{n^2}{\pi^2}\Big{[}\sum_{k=1}^d(n_kR)^2+\Big{(}\frac{n{\,}(L-a)}{\pi}\Big{)}^2\Big{]}^{\frac{1}{2}(s-\frac{D-1}{2}-\frac{d+1}{2}-\frac{3}{2})}\times
\\& \notag K_{1+\frac{d+1}{2}-(s-\frac{D-1}{2})}\bigg{(}2\pi
\sqrt{J(J+1)+m^2}{\,}\Big{[}\sum_{k=1}^d(n_kR)^2+\Big{(}\frac{n{\,}(L-a)}{\pi}\Big{)}^2\Big{]}^{\frac{1}{2}}\bigg{)}\Bigg{\}}
\end{align}
The above relation is not so ''friendly''. However within
approximation $a\gg R$ and $L-a\gg R$, the argument of the Bessel
functions above are large. Thus we can use the asymptotic
expansion,
\begin{equation}\label{besselappr}
K_{\nu}(z)=\sqrt{\frac{\pi}{2z}}e^{-z}\Big{(}1+\frac{\nu-1}{8z}+....\Big{)}.
\end{equation}
It is obvious that the leading contributions comes when keeping
only the $n_i=1$ terms in the summations. So finally the Casimir
force equals to,
\begin{align}\label{aftercasimirrokies}
&F_c(s)=-\sum_{J=0}^N\frac{1}{4}\pi^{s-2}{\,}R^d\frac{\Big{(}\sqrt{J(J+1)+m^2}{\,}\Big{)}^{\frac{d+1}{2}-(s-\frac{D-1}{2}-\frac{1}{2})}}{\Gamma(s)}
\times \\& \notag \Bigg{\{}
\frac{1}{2}\Big{[}dR^2+\frac{a^2}{\pi^2}\Big{]}^{\frac{1}{2}(s-\frac{D-1}{2}-\frac{d+1}{2}-2)}e^{-2\pi(J(J+1)+m^2)\sqrt{dR^2+\frac{a^2}{\pi^2}}}
\\& \notag
-\frac{1}{2}\Big{[}dR^2+\frac{(L-a)^2}{\pi^2}\Big{]}^{\frac{1}{2}(s-\frac{D-1}{2}-\frac{d+1}{2}-2)}e^{-2\pi(J(J+1)+m^2)\sqrt{dR^2+\frac{(L-a)^2}{\pi^2}}}
\\& \notag -a^2{\,}\Big{(}\frac{d+1}{2}-(s-\frac{D-1}{2})\Big{)}\Big{(}\sqrt{J(J+1)+m^2}{\,}\Big{)}\times \\& \notag
\frac{1}{2\pi^2}\Big{[}dR^2+\frac{a^2}{\pi^2}\Big{]}^{\frac{1}{2}(s-\frac{D-1}{2}-\frac{d+1}{2}-2)}e^{-2\pi(J(J+1)+m^2)\sqrt{dR^2+\frac{a^2}{\pi^2}}}
\\& \notag+(L-a)^2{\,}\Big{(}\frac{d+1}{2}-(s-\frac{D-1}{2})\Big{)}\times
\\& \notag
\frac{1}{2\pi^2}\Big{(}\sqrt{J(J+1)+m^2}\Big{)}\Big{[}dR^2+\frac{(L-a)^2}{\pi^2}\Big{]}^{\frac{1}{2}(s-\frac{D-1}{2}-\frac{d+1}{2}-2)}e^{-2\pi(J(J+1)+m^2)\sqrt{dR^2+\frac{(L-a)^2}{\pi^2}}}\Bigg{\}}
\end{align}
Let us now discuss the behavior of the Casimir force as a function
of $a$. First notice that the parameter $R^d$ in the first line in
combination with the exponentials make the Casimir force really
small in magnitude. This is to be contrasted to the case when only
the fuzzy sphere is the extra dimensional space.

\noindent Now in order to be correct with the values that $a$ is
allowed to take we mention that (\ref{aftercasimirrokies}) is
obtained in the limit $a\gg 1$ and $L-a\gg 1$. Thus the extremely
small values for $a$ are excluded. So we shall examine the values
of $a$ that change the sign of $L-a$, that is $a<\frac{L}{2}$ and
$a>\frac{L}{2}$. It is obvious that when $a\gg 1$ and $L-a\gg 1$
and $a<\frac{L}{2}$ the most dominating terms in relation
(\ref{aftercasimirrokies}) are the last two terms. For $D=3$ and
$s=-\frac{1}{2}$ and for $a<\frac{L}{2}$, the Casimir force is
negative because the first of the two last terms is dominating. So
as in the cases we mentioned in the introduction, the piston is
attracted to the closest end. Now when $a>\frac{L}{2}$ the most
dominating term of the last two is the second one. Thus the
Casimir force is positive. So the Casimir force is repulsive in
this case.

\noindent Before ending this section let us discuss on a topic.
Notice that the last two terms of relation
(\ref{aftercasimirrokies}) are multiplied by $\sqrt{J(J+1)+m^2}$.
When the fuzzy sphere is excluded from the calculations, one can
say that due to this term in the small mass limit the last two
terms cease to be dominating. So the Casimir force changes sign.
However in our case this does not happens because the finite
values of order $\sim 1$ that $J(J+1)$ can take. Thus even in the
small mass limit, the sign of the Casimir force does not change
from the behavior we described previously.

\section{Casimir piston with a fuzzy sphere extra dimension}

We now consider a fuzzy sphere as the extra dimensional space. We
again deal the case with general $s$ and $D$. In order to get the
case of our interest and study the singularities, at the end we
put $s=-\frac{1}{2}$ and $D=3$, as previously. The Casimir energy
in this case is,
\begin{align}
E_c(s,a)=\frac{1}{4\pi^2}\int
\mathrm{d}^{D-1}p\sum_{J=0}^N\sum_{n=1}^{\infty}\Big{[}\sum_{k=1}^{D-1}p_k^2+(\frac{n\pi}{a})^2+J(J+1)+m^2\Big{]}^{-s}.
\end{align}
which upon integrating over the continuous modes and using
(\ref{aplodiplo}), we obtain,
\begin{align}\label{pord12}
&E_c(s,a)=\frac{1}{8\pi^2}\pi^{\frac{D-1}{2}}\frac{\Gamma(s-\frac{D-1}{2})}{\Gamma(s)}
\times\\& \notag
\Big{(}\sum_{J=0}^N\sum_{n=-\infty}^{\infty}\Big{[}(\frac{n\pi}{a})^2+J(J+1)+m^2\Big{]}^{\frac{D-1}{2}-s}
-\sum_{J=0}^N\Big{[}J(J+1)+m^2\Big{]}^{\frac{D-1}{2}-s}\Big{)}
\end{align}
Upon using as before relation (\ref{cativoepsteinzeta}),
\begin{align}\label{pord12}
&E_c(s,a)=\frac{1}{8\pi^2}\pi^{\frac{D-1}{2}}\frac{\Gamma(s-\frac{D-1}{2})}{\Gamma(s)}
\times\\& \notag
\Big{(}\sum_{J=0}^NZ_{1}^{w^2_1}(s-\frac{D-1}{2}{\,};\frac{\pi^2}{a^2})
-\sum_{J=0}^N\Big{[}J(J+1)+m^2\Big{]}^{\frac{D-1}{2}-s}\Big{)}
\end{align}
Note that $w^2_1=J(J+1)+m^2$, as previously. Using (\ref{kepst1}),
relation (\ref{pord12}) becomes,
\begin{align}\label{camew1}
&E_c(s,a)=\frac{1}{8\pi^2}\pi^{\frac{D-1}{2}}\frac{\Gamma(s-\frac{D-1}{2})}{\Gamma(s)}
\times \\& \notag
\sum_{J=0}^N\Bigg{\{}\frac{a{\,}\pi^{-\frac{1}{2}}{\,}\Gamma(s-\frac{D-1}{2}-\frac{1}{2})}{\Gamma(s-\frac{D-1}{2})}{\,}\Big{(}\sqrt{J(J+1)+m^2}{\,}\Big{)}^{1-2(s-\frac{D-1}{2})}\\&
\notag+\frac{2\pi^{s-\frac{D}{2}-1}{\,}a{\,}\Big{(}\sqrt{J(J+1)+m^2}{\,}\Big{)}^{\frac{1}{2}-(s-\frac{D-1}{2})}}{\Gamma(s-\frac{D-1}{2})}
\times \\& \notag
\sum^{\infty'}_{n=-\infty}\Big{[}\frac{n{\,}a}{\pi}\Big{]}^{(s-\frac{D-1}{2}-\frac{1}{2})}
 K_{\frac{1}{2}-(s-\frac{D-1}{2})}\bigg{(}2n{\,}a\sqrt{J(J+1)+m^2}{\,}\bigg{)}\\& \notag-\Big{[}J(J+1)+m^2\Big{]}^{\frac{D-1}{2}-s}\Bigg{\}}
\end{align}
It is obvious that due to the gamma function
$\Gamma(s-\frac{D-1}{2}-\frac{1}{2})$, for the values
$s=-\frac{1}{2}$ and $D=3$, a singularity appears in the first
term of the above expression. However when we compute the Casimir
force, as in the previous section, the linearity of this term to
$a$ and the addition of the other piston chamber will cancel this
singular term. Following the procedure of the previous section we
obtain the Casimir force (we use the limit $L-a\gg 1$ and $a\gg
1$),
\begin{align}\label{aftercaspoutsies}
&F_c(s)=-\sum_{J=0}^N\frac{1}{4}\pi^{s-2}{\,}\frac{\Big{(}\sqrt{J(J+1)+m^2}{\,}\Big{)}^{\frac{1}{2}-(s-\frac{D-1}{2}-\frac{1}{2})}}{\Gamma(s)}
\times \\& \notag \Bigg{\{}
\frac{1}{2}\Big{[}\frac{a}{\pi}\Big{]}^{(s-\frac{D-1}{2}-\frac{1}{2}-2)}e^{-2\pi(J(J+1)+m^2)\frac{a}{\pi}}
\\& \notag
-\frac{1}{2}\Big{[}\frac{L-a}{\pi}\Big{]}^{(s-\frac{D-1}{2}-\frac{1}{2}-2)}e^{-2\pi(J(J+1)+m^2)\frac{(L-a)}{\pi}}
\\& \notag -a^2{\,}\Big{(}\frac{1}{2}-(s-\frac{D-1}{2})\Big{)}\Big{(}\sqrt{J(J+1)+m^2}{\,}\Big{)}\times \\& \notag
\frac{1}{2\pi^2}\Big{[}\frac{a}{\pi}\Big{]}^{(s-\frac{D-1}{2}-\frac{1}{2}-2)}e^{-2\pi(J(J+1)+m^2)\frac{a}{\pi}}
\\& \notag+(L-a)^2{\,}\Big{(}\frac{1}{2}-(s-\frac{D-1}{2})\Big{)}\times
\\& \notag
\frac{1}{2\pi^2}\Big{(}\sqrt{J(J+1)+m^2}\Big{)}\Big{[}\frac{L-a}{\pi}\Big{]}^{(s-\frac{D-1}{2}-\frac{1}{2}-2)}e^{-2\pi(J(J+1)+m^2)\frac{(L-a)}{\pi}}\Bigg{\}}
\end{align}
As in the previous section case, the same arguments hold for the
allowed values of $a$. Thus when $D=3$ and $s=-\frac{1}{2}$ and
for $a<\frac{L}{2}$ and $L-a\gg 1$ and $a\gg 1$, the two terms of
relation (\ref{aftercaspoutsies}) dominate making the Casimir
force negative. Thus the piston is attracted to the nearest end.
However when $a>\frac{L}{2}$ the force is positive and thus
repulsive. So the behavior of the sign is the same as in the
previous section case where an extra dimensional commutative torus
and the fuzzy sphere was the extra dimensional space.

\noindent Note that the difference with the torus case of the
previous section is that the Casimir force is not as small was the
force for $T^d\times S_{FZ}$, because the $R^d$ term does not
appear here. However the force is exponentially suppressed as in
the previous section.

\section{Computation of Casimir energy adding one loop self energy corrections}

We discussed in the introduction the motivation to use
non-commutative extra dimensions. The need comes from the
stabilization they offer to the commutative extra dimensions. A
calculation that follows naturally in this setup is the
calculation of the Casimir energy. Also great interest comes from
the computation of the Casimir energy by adding one loop
corrections to the self energy of the scalar field due to
non-commutative extra dimension. This was computed in reference
\cite{huang} for the case $M^D\times T^d\times S_{FZ}$. We present
a correct brief calculation in the appendix. $M^D$ represents the
$D$-dimensional Minkowski spacetime. The inclusion of higher loop
corrections due to self interactions of scalar field in a
spacetime with non-commutative extra dimensions is frequently done
in such theories (see \cite{huang,nam,myers}). However in the case
of interest a problem arises when $D$=even. The Casimir energy is
computed with the approximation $\frac{1}{R^2}\gg m^2$ and
$\frac{1}{R^2}\gg N(N+1)$. Within this approximation the one loop
correction to the Casimir energy, as we prove in detail in the
appendix, when $D=$even, it contains singularities. Also when
$d=$odd the Casimir energy also contains singularities. Indeed the
Casimir energy reads,
\begin{align}\label{ezetaiddfnbefore}
&E_c(s)=\frac{1}{4\pi^2}\pi^{\frac{D-1}{2}}\frac{\Gamma(s-\frac{D-1}{2})}{\Gamma(s)}\times
\\& \notag
\Big{\{}\Big{[}\frac{R^d\pi^{\frac{d}{2}}}{(2\pi)^d}\frac{\Gamma(s-\frac{D-1}{2}-\frac{d}{2})}{\Gamma(s-\frac{D-1}{2})}{\,}w^{d-2(s-\frac{D-1}{2})}\\&
\notag+\frac{2\pi^{s-\frac{D-1}{2}}w^{d-2(s-\frac{D-1}{2})}}{\Gamma(s-\frac{D-1}{2})}\sum^{\infty'}_{n_1,n_2,...,{\,}n_d=-\infty}\Big{[}\sum_{k=1}^d\Big{(}\frac{n_kR}{2\pi}\Big{)}^2\Big{]}^{\frac{1}{2}(s-\frac{D-1}{2}-\frac{d}{2})}\times
\\& \notag K_{\frac{d}{2}-(s-\frac{D-1}{2})}\Big{(}2\pi w\Big{[}\sum_{k=1}^d\big{(}\frac{n_kR}{2\pi}\big{)}^2\Big{]}^{\frac{1}{2}}\Big{)}-(AR^{2-D})^{\frac{D-1}{2}-s}\Big{\}},
\end{align}
where $A$ stands for,
\begin{equation}\label{pesimoxesimo}
A=\frac{g}{24}\pi^{\frac{D}{2}}(3N^2+6N+1)Z_d(2-D)\Gamma(1-\frac{D}{2})(\frac{1}{2\pi})^{2-D}.
\end{equation}
In the above relations the problems are clearly seen (note that
$s=-\frac{1}{2}$ in the end of the calculation). Thus the most
phenomenologically interesting case, $D=4$, is out of the
question. This motivated us to calculate the Casimir energy in the
case of two parallel plates separated at a distance $a$ and maybe
generalize the calculation to a piston setup to see if the usual
regularization the Casimir piston offers to the force, hold here
too (this is because the dimensionality of the infinite dimensions
reduces to three). We computed the one loop corrections for the
cases of $S_{FZ}$ and $S_{FZ}\times T^d$ as extra dimensional
space. In the case with $S_{FZ}\times T^d$, the most easy and most
interesting case, from a computational point of view, arises when
$R=\frac{a}{\pi}$. All the calculations are done within the
approximations, $\frac{1}{R^2}\gg m^2$ and $\frac{1}{R^2}\gg
N(N+1)$.

\subsection{One loop corrections to the Casimir energy of a piston chamber with $S_{FZ}$ as extra dimensional space}

We shall consider firstly the case where the extra dimensional
space is only the fuzzy sphere. We deal this case first because it
is simpler.

\noindent With the incorporation of the 1-loop self energy
corrections due to self interactions in non-commutative geometry,
the Casimir energy for the piston chamber reads,
\begin{align}
E_c(s,a)=\frac{1}{4\pi^2}\int
\mathrm{d}^{D-1}p\sum_{J=0}^N\sum_{n=1}^{\infty}\Big{[}\sum_{k=1}^{D-1}p_k^2+(\frac{n\pi}{a})^2+J(J+1)+m^2+\Sigma\Big{]}^{-s}.
\end{align}
In the above expression, $\Sigma$ is the one loop corrections and
equals to,
\begin{equation}\label{sig}
\Sigma=\Gamma_{\mathrm{planar}}^{(2)}+\Gamma_{\mathrm{nonplanar}}^{(2)}
\end{equation}
with,
\begin{equation}\label{plw}
\Gamma_{\mathrm{planar}}^{(2)}=\frac{g}{12\pi}I^P,
\end{equation}
and,
\begin{equation}\label{nonplw}
\Gamma_{\mathrm{nonplanar}}^{(2)}=\frac{g}{24\pi}I^{NP}.
\end{equation}
$I^P$ and $I^{NP}$ are the planar and non-planar contributions
respectively and are, (we follow references \cite{nam} and
\cite{huang}),
\begin{equation}\label{planar}
I^P=\int\frac{\mathrm{d}^{D}\vec{p}}{(2\pi)^{D}}\sum_{n=1}^{\infty}\sum_{J=0}^N\frac{2J+1}{\vec{p}^{{\,}2}+(n\pi/a)^2+J(J+1)+m^2}
\end{equation}
and the non-planar contribution,
\begin{eqnarray}\label{dynamic1}
I^{NP}&=&\int\frac{\mathrm{d}^{D}\vec{p}}{(2\pi)^{D}}\sum_{n=1}^{\infty}\sum_{J=0}^N(-1)^{J+j+N}
\\ && \nonumber \frac{(2J+1)(N+1)}{\vec{p}^{{\,}2}+(n\pi/a)^2+J(J+1)+m^2}{\,}\bigg{\{}
\begin{array}{ccc}
  N/2 & N/2 & j \\
  N/2 & N/2 & J \\
\end{array}\bigg{\}}.
\end{eqnarray}
The index $j$ is to be contracted with the symbol in curly
brackets. This symbol corresponds to the Wigner symbol of the
$SU(2)$ algebra, see \cite{huang}. We now compute $I^P$ and
$I^{NP}$. We will need the following properties of the Wigner
symbol,
\begin{equation}\label{wigner1}
\sum_N(2N+1)\bigg{\{}\begin{array}{ccc}
  A & B & N \\
  C & D & P \\
\end{array}\bigg{\}}
\bigg{\{}\begin{array}{ccc}
  A & B & N \\
  C & D & Q \\
\end{array}\bigg{\}}=\frac{1}{2P+1}\delta_{PQ}
\end{equation}
and also,
\begin{equation}\label{wigner2}
\sum_N(-1)^{N+P+Q}(2N+1)\bigg{\{}\begin{array}{ccc}
  A & B & N \\
  C & D & P \\
\end{array}\bigg{\}}
\bigg{\{}\begin{array}{ccc}
  A & B & N \\
  D & C & Q \\
\end{array}\bigg{\}}=\begin{array}{ccc}
  A & C & Q \\
  B & C & P \\
\end{array}\bigg{\}}
\end{equation}
An interesting computational case arises in the approximation,
\begin{equation}\label{appr1}
\frac{\pi^2}{a^2}\gg N(N+1),{\,}{\,}{\,}{\,}\frac{\pi^2}{a^2}\gg
m^2
\end{equation}
This approximation is applicable in the case when the plate
separation is very small in magnitude. Note that within this
approximation we cannot take the $N\rightarrow \infty$ limit,
which corresponds to the continuum two dimensional commutative
sphere \cite{huang}. With approximation (\ref{appr1}), the planar
contribution becomes,
\begin{equation}\label{pc1}
I^P=\pi^{\frac{D}{2}}a^{2-D}\Gamma(1-\frac{D}{2})\sum_{n=1}^{\infty}\sum_{J=0}^N\frac{2J+1}{(n^2\pi^2)^{1-\frac{D}{2}}}
\end{equation}
Remember that in the end we take $D=3$ again. The integration over
momenta differs on the initial calculation of the Casimir energy.
This is because the loop corrections include total spacetime
integration and not just space integration (in all the cases the
integrations are assumed to be Euclidean). So the planar and
non-planar contributions (see relations (\ref{planar}) and
(\ref{dynamic1})) are integrated over $D$ continuous dimensions
which are, $D-1$ space dimensions and 1 time dimension. Using the
Riemann zeta function, the planar contribution becomes,
\begin{equation}\label{pc2}
I^P=\pi^{\frac{D}{2}}a^{2-D}\Gamma(1-\frac{D}{2})(N^2+2N)\zeta(2-D)
\end{equation}
In the same way the non planar contribution reads,
\begin{equation}\label{cp34}
I^{NP}=\pi^{\frac{3D}{2}-2}a^{2-D}\Gamma(1-\frac{D}{2})(N+1)^2\zeta(2-D)
\end{equation}
Thus relation (\ref{sig}) becomes,
\begin{equation}\label{desig}
\Sigma=\frac{g}{24}\pi^{\frac{3D}{2}-2}a^{2-D}\Gamma(1-\frac{D}{2})(3N^2+6N+1)\zeta(2-D)
\end{equation}
and in more compact form
\begin{equation}\label{frgr}
\Sigma=\sigma_1a^{2-D}
\end{equation}
and
$\sigma=\frac{g}{24}\pi^{\frac{3D}{2}-2}\Gamma(1-\frac{D}{2})(3N^2+6N+1)\zeta(2-D)$.
Within the approximation (\ref{appr1}), the Casimir energy reads,
\begin{equation}\label{dse}
E_c(s,a)=\frac{1}{8\pi^2}\pi^{\frac{D-1}{2}}\frac{\Gamma(s-\frac{D-1}{2})}{\Gamma(s)}\sum_{n=1}^{\infty}\Big{[}(\frac{n\pi}{a})^2+\sigma_1a^{2-D}\Big{]}^{\frac{D-1}{2}-s}
\end{equation}
As in previous, using (\ref{kepst1}), (\ref{aplodiplo}),
\begin{align}\label{camew}
&E_c(s,a)=\frac{1}{8\pi^2}\pi^{\frac{D-1}{2}}\frac{\Gamma(s-\frac{D-1}{2})}{\Gamma(s)}
\times \\& \notag
\Bigg{\{}\frac{a{\,}\pi^{-\frac{1}{2}}{\,}\Gamma(s-\frac{D-1}{2}-\frac{1}{2})}{\Gamma(s-\frac{D-1}{2})}{\,}\Big{(}\sqrt{\sigma_1a^{2-D}}{\,}\Big{)}^{1-2(s-\frac{D-1}{2})}\\&
\notag+\frac{2\pi^{s-\frac{D}{2}-1}{\,}a{\,}\Big{(}\sqrt{\sigma_1a^{2-D}}{\,}\Big{)}^{\frac{1}{2}-(s-\frac{D-1}{2})}}{\Gamma(s-\frac{D-1}{2})}
\times \\& \notag
\sum^{\infty'}_{n=-\infty}\Big{[}\frac{n{\,}a}{\pi}\Big{]}^{(s-\frac{D-1}{2}-\frac{1}{2})}
 K_{\frac{1}{2}-(s-\frac{D-1}{2})}\bigg{(}2n{\,}a\sqrt{\sigma_1a^{2-D}}{\,}\bigg{)}\\& \notag-\Big{[}\sigma_1a^{2-D}\Big{]}^{\frac{D-1}{2}-s}\Bigg{\}}
\end{align}
The last relation clearly diverges for $D=3$ and $s=-\frac{1}{2}$.
Unfortunately computing the Casimir force and adding the
contributions from the two chambers still does not yield a regular
expression. This is because the linearity of the first term of
(\ref{camew}) is lost.

\subsection{One loop Casimir energy of a piston chamber with $T^d\times S_{FZ}$ as extra dimensional space}

We now compute the 1-loop self energy corrected Casimir energy for
the $a$ piston chamber in the case the extra dimensional space is
$T^d\times S_{FZ}$. The most easy and interesting case is when,
\begin{equation}\label{skat}
R=\frac{a}{\pi}
\end{equation}
We now compute the planar and non planar graphs within this
approximation. The planar contribution is,
\begin{equation}\label{planar12}
I^P=\int\frac{\mathrm{d}^{D-1}p}{(2\pi)^{D-1}}\sum_{n=1}^{\infty}\sum^{\infty}_{n_1,{\,}n_2,...,{\,}n_{d}=-\infty}\sum_{J=0}^N\frac{2J+1}{\sum_{k=1}^{D-1}p_k^2+(n\pi/a)^2+(2\pi
\vec{n}/R)^2+J(J+1)+m^2}
\end{equation}
Within the approximation (\ref{appr1}), and using the same
techniques as in previous we obtain,
\begin{equation}\label{dfra}
I^P=\pi^{\frac{3D}{2}-2}\Gamma(1-\frac{D}{2})(N^2+2N)\bigg{(}\zeta(2-D)+\frac{1}{2}Z_{1+d}(1-\frac{D}{2})\bigg{)}a^{2-D}
\end{equation}
In the same way, the non-planar contribution reads,
\begin{equation}\label{fghudf}
I^{NP}=\pi^{\frac{3D}{2}-2}\Gamma(1-\frac{D}{2})(N+1)^2\bigg{(}\zeta(2-D)+\frac{1}{2}Z_{1+d}(1-\frac{D}{2})\bigg{)}a^{2-D}
\end{equation}
So the one loop correction to the self energy is,
\begin{equation}\label{ghtj}
\Sigma=\sigma_2a^{2-D}
\end{equation}
with,
\begin{equation}\label{utgyurt}
\sigma_2=\frac{g}{24}\pi^{\frac{3D}{2}-2}\Gamma(1-\frac{D}{2})(3N^2+6N+1)^2\bigg{(}\zeta(2-D)+\frac{1}{2}Z_{1+d}(1-\frac{D}{2})\bigg{)}
\end{equation}
Now the one loop self energy corrected Casimir energy in this case
reads,
\begin{align}
E_c(s,a)=\frac{1}{4\pi^2}\int
\mathrm{d}^{D-1}p\sum_{J=0}^N\sum_{n=1}^{\infty}\sum^{\infty}_{n_1,n_2,...,{\,}n_d=-\infty}\Big{[}\sum_{k=1}^{D-1}p_k^2+(\frac{n\pi}{a})^2+\sum_{k=1}^d(\frac{n_k\pi}{a})^2+\Sigma\Big{]}^{-s}.
\end{align}
Following the steps of the previous sections, we obtain finally,
\begin{align}\label{po671334}
&E_c(s,a)=\frac{1}{8\pi^2}\pi^{\frac{D-1}{2}}\frac{\Gamma(s-\frac{D-1}{2})}{\Gamma(s)}
\times \\& \notag
\Bigg{\{}\frac{a^{d+1}{\,}\pi^{d}{\,}\Gamma(s-\frac{D-1}{2}-\frac{d+1}{2})}{\pi^{d+1}\Gamma(s-\frac{D-1}{2})}{\,}\Big{(}\sqrt{\sigma_2a^{2-D}}{\,}\Big{)}^{d+1-2(s-\frac{D-1}{2})}\\&
\notag+\frac{2\pi^{s-\frac{D-1}{2}}{\,}a^{d+1}{\,}\Big{(}\sqrt{\sigma_2a^{2-D}}{\,}\Big{)}^{\frac{d+1}{2}-(s-\frac{D-1}{2})}}{\pi^{d+1}\Gamma(s-\frac{D-1}{2})}
\times \\& \notag
\sum^{\infty'}_{n,n_1,n_2,...,{\,}n_d=-\infty}\Big{[}\sum_{k=1}^{d+1}\Big{(}\frac{n_ka}{\pi}\Big{)}^2\Big{]}^{\frac{1}{2}(s-\frac{D-1}{2}-\frac{d+1}{2})}\times
\\& \notag K_{\frac{d+1}{2}-(s-\frac{D-1}{2})}\bigg{(}2\pi \sqrt{\sigma_2a^{2-D}}{\,}\Big{[}\sum_{k=1}^{d+1}\Big{(}\frac{n_k{\,}a}{\pi}\Big{)}^2\Big{]}^{\frac{1}{2}}\bigg{)}\\& \notag
-\frac{a^{d}{\,}\pi^{d/2}\Gamma(s-\frac{D-1}{2}-\frac{d}{2})}{\Gamma(s-\frac{D-1}{2})}{\,}\Big{(}\sqrt{\sigma_2a^{2-D}}{\,}\Big{)}^{d-2(s-\frac{D-1}{2})}\\&
\notag-\frac{2\pi^{s-\frac{D-1}{2}}a^d\Big{(}\sqrt{\sigma_2a^{2-D}}{\,}\Big{)}^{\frac{d}{2}-(s-\frac{D-1}{2})}}{\Gamma(s-\frac{D-1}{2})}\sum^{\infty'}_{n_1,n_2,...,{\,}n_d=-\infty}\Big{[}\sum_{k=1}^d\Big{(}\frac{n_ka}{\pi}\Big{)}^2\Big{]}^{\frac{1}{2}(s-\frac{D-1}{2}-\frac{d}{2})}\times
\\& \notag K_{\frac{d}{2}-(s-\frac{D-1}{2})}\bigg{(}2\pi \sqrt{\sigma_2a^{2-D}}{\,}\Big{[}\sum_{k=1}^d\Big{(}\frac{n_ka}{\pi}\Big{)}^2\Big{]}^{\frac{1}{2}}\bigg{)}\Bigg{\}}
\end{align}
It is obvious that for $s=-\frac{1}{2}$ and $D=3$ relation
(\ref{po671334}) contains singularities. Actually the
singularities are contained in the gamma functions
$\Gamma(s-\frac{D-1}{2}-\frac{d+1}{2})$ and
$\Gamma(s-\frac{D-1}{2}-\frac{d}{2})$ for $d$ even and odd.
Unfortunately the Casimir energy is divergent for all $D$ values.
Also the Casimir force is also divergent for all $d$ values. We
expected the piston configuration would lead to a regularization
of the Casimir force which is ill defined in the case the plates
are excluded. However even in the piston case this does not
happens because the singularities of the Casimir energy are not
linear to $a$ as happened in the previous sections, see for
example relations (\ref{camew}) and (\ref{camew1}) and no
cancellation due to piston setup occurs.

\noindent Thus there is no consistent way to incorporate the one
loop corrections to the scalar mass within the approximation
(\ref{skat}) and (\ref{appr1}) even in the piston setup. Maybe the
problem is the approximations or the dimensionality of spacetime,
as in the Minkowski case (see appendix).

\section{Conclusions}

We have examined how the Casimir force for a piston behaves when
non-commutative extra dimensions are included in the calculations.
Particularly we have examined the cases when the extra dimensional
space is $T^d\times S_{FZ}$ and $S_{FZ}$, with $T^d$ the
commutative $d$ torus and $S_{FZ}$ the two dimensional
non-commutative sphere. For the case $T^d\times S_{FZ}$, when the
two piston chambers are much larger than the compact torus radius,
the Casimir force behaves as follows,

\begin{itemize}

\item When $a>\frac{L}{2}$ the force is repulsive.

\item When $a<\frac{L}{2}$ the force is attractive and the piston
is attracted to the nearest end.

\end{itemize}

\noindent In the case of the fuzzy sphere as extra dimensional
space the results are the same. Also we concluded that the small
mass limit will not affect our results due to the
non-commutativity effects. Thus this behavior for $a\gg 1$ and
$L-a\gg 1$ remain true for all the dimensions of the extra
dimensional torus. Additionally the force in the case $T^d\times
S_{FZ}$ is very suppressed due to the $R^d$ multiplying the whole
expression. This suppression does not appear in the case of
$S_{FZ}$. We also examined the incorporation of one loop
corrections to the self energy due to the self interactions of the
scalar field in the non-commutative space. The most mathematically
interesting case was when $R=\frac{a}{\pi}$. We did this for one
piston chamber. We expected to obtain a regular expression for the
Casimir force. Without the plate separation the Casimir energy was
singular for $D=4$ and generally for $D$=even. Thus we hoped that
the introduction of plates would solve the problem. However this
did not happened, but contrary the result is completely singular.
Even the addition of the other chamber of the piston did not
change the result. We proved that this happens when the ''mass''
of the scalar field is $a$ dependent. We have tried also Neumann
boundary conditions but the results did not change.

\noindent Before closing we discuss on the stabilization issue.
The non-commutative space having intrinsically a scale, the Moyal
shell, one expects to stabilize the commutative extra dimensions
(of course if only non-commutative dimensions appear they are by
definition stabilized). Thus we should check on this for the
piston setup. This could be studied if we computed the force with
respect to $R$,
\begin{equation}\label{stabiliz}
F_R=-\frac{\partial E_c}{\partial R}.
\end{equation}
Unfortunately the expression contains singularities, as we can see
by taking the derivatives over $R$ of relations
(\ref{pordoulis1334}) and (\ref{camew1}). Also this remains true
when only the commutative torus is the only extra dimensional
space. Thus we must find a regularization method in order to
answer this problem. We hope to comment on this soon.

\bigskip
\bigskip

\section*{APPENDIX}

We will now compute the Casimir energy in the case the extra
dimensions are $T^d\times S_{FZ}$ and for the four dimensional
spacetime (for the one loop corrections see \cite{huang}). We
shall use the approximation,
\begin{equation}\label{apdddd2}
\frac{\pi^2}{R^2}\gg N(N+1),{\,}{\,}{\,}{\,}\frac{\pi^2}{R^2}\gg
m^2.
\end{equation}
The one loop corrected Casimir energy is equal to,
\begin{equation}\label{casimir2we}
E_c(s)=\int
\mathrm{d}^{D-1}p\sum^{\infty'}_{n_1,n_2,...,{\,}n_d=-\infty}\sum_{J=0}^N\Big{[}\sum_{k=1}^d\big{(}\frac{
n_k}{R}\big{)}^2+\vec{p}{\,}^2+J(J+1)+m^2+\Sigma\Big{]}^{-s}.
\end{equation}
In this case at the end of the calculation, $s=-\frac{1}{2}$ and
$D=4$ contrary to the piston case where $D=3$. $\Sigma$ equals to
the self energy 1-loop correction and as previously equals to,
\begin{equation}\label{ed}
\Sigma=\frac{g}{12\pi}I^P+\frac{g}{24\pi}I^{NP}.
\end{equation}
As before,
\begin{equation}\label{bersalio}
I^P=\pi^{\frac{D}{2}}\Gamma(1-\frac{D}{2})\sum^{\infty'}_{n_1,n_2,...,{\,}n_d=-\infty}\sum_{J=0}^{N}\frac{2J+1}{\sum_{k=1}^d(\frac{
n_k}{R})^2+J(J+1)+m^2}.
\end{equation}
Note that in this case the approximation (\ref{apdddd2}) is valid
if the zero modes of the commutative extra dimensions are excluded
for the summation, contrary to the piston case. After
calculations, $I^P$ equals to,
\begin{equation}\label{neddrip}
I^P=\pi^{\frac{D}{2}}(\frac{R}{2\pi})^{2-D}(N^2+2N)Z_d(2-D)\Gamma(1-\frac{D}{2})
\end{equation}
and in the same way, $I^{NP}$,
\begin{equation}\label{inp}
I^{NP}=\pi^{\frac{D}{2}}(\frac{R}{2\pi})^{2-D}(N+1)^2Z_d(2-D)\Gamma(1-\frac{D}{2}).
\end{equation}
Finally,
\begin{equation}\label{sigma}
\Sigma=AR^{2-D},
\end{equation}
with,
\begin{equation}\label{pesimo}
A=\frac{g}{24}\pi^{\frac{D}{2}}(3N^2+6N+1)Z_d(2-D)\Gamma(1-\frac{D}{2})(\frac{1}{2\pi})^{2-D}.
\end{equation}
For more details see \cite{huang}. Now the Casimir energy, after
integrating over the continuous $D-1$ dimensions (that is 3
dimensions),
\begin{equation}\label{intdfpq}
E_c(s)=\frac{1}{4\pi^2}\pi^{\frac{D-1}{2}}\frac{\Gamma(s-\frac{D-1}{2})}{\Gamma(s)}\sum^{\infty'}_{n_1,n_2,...,{\,}n_d=-\infty}\Big{[}\sum_{k=1}^d\big{(}\frac{2\pi
n_k}{R}\big{)}^2+AR^{2-D}\Big{]}^{\frac{D-1}{2}-s}.
\end{equation}
Including the zero mode in the sum and using the homogeneous
Epstein zeta, the above becomes,
\begin{align}\label{ezetaiddfn}
&E_c(s)=\frac{1}{4\pi^2}\pi^{\frac{D-1}{2}}\frac{\Gamma(s-\frac{D-1}{2})}{\Gamma(s)}\times
\\& \notag
\Big{\{}\Big{[}\frac{R^d\pi^{\frac{d}{2}}}{(2\pi)^d}\frac{\Gamma(s-\frac{D-1}{2}-\frac{d}{2})}{\Gamma(s-\frac{D-1}{2})}{\,}w^{d-2(s-\frac{D-1}{2})}\\&
\notag+\frac{2\pi^{s-\frac{D-1}{2}}w^{d-2(s-\frac{D-1}{2})}}{\Gamma(s-\frac{D-1}{2})}\sum^{\infty'}_{n_1,n_2,...,{\,}n_d=-\infty}\Big{[}\sum_{k=1}^d\Big{(}\frac{n_kR}{2\pi}\Big{)}^2\Big{]}^{\frac{1}{2}(s-\frac{D-1}{2}-\frac{d}{2})}\times
\\& \notag K_{\frac{d}{2}-(s-\frac{D-1}{2})}\Big{(}2\pi w\Big{[}\sum_{k=1}^d\big{(}\frac{n_kR}{2\pi}\big{)}^2\Big{]}^{\frac{1}{2}}\Big{)}-(AR^{2-D})^{\frac{D-1}{2}-s}\Big{\}},
\end{align}
and $w^2=AR^{2-D}$. From relation (\ref{pesimo}) and
(\ref{ezetaiddfn}) it is obvious that the only singularity free
case is when $D=$odd and $d=$even. Unfortunately this does not
correspond to our world because $D=4$. This is why we did the
calculation for the piston configuration as we pointed in previous
sections. In the piston setup singularities don't appear (at least
in the force) and the continuous dimensions are reduced by one. So
we could expect that the ill calculation we just made could be
done. Unfortunately this was not true as we saw in the previous
sections.


\begin{thebibliography}{99}



\bibitem{Casimir} H. Casimir, Proc. Kon. Nederl. Akad. Wet. 51 793
(1948)

\bibitem{KirstenKaluzaKleinPiston} K. Kirsten, S. A. Fulling,
\arx0901.1902

\bibitem{Cheng} H. Cheng, \plb 668, 72 (2008)

\bibitem{huang}Wung-Hong Huang, Phys. Lett. B537, 311 (2002)


\bibitem{huang1} Wung-Hong Huang, JHEP 0011, 041 (2000)

\bibitem{huang2} Wung-Hong Huang, JHEP 0207, 064 (2002)

\bibitem{huang3} Wung-Hong Huang, arXiv:0901.0614 [hep-th]

\bibitem{elizalde1} A.A. Bytsenko, E. Elizalde, S. Zerbini, Phys. Rev. D64, 105024
(2001)

\bibitem{myers}Robert C. Myers, JHEP 9912, 022 (1999)

\bibitem{gaumis} Jaume Gomis, Thomas Mehen, Mark B. Wise, JHEP 0008, 029
(2000)

\bibitem{tsexoi} H. Grosse, C. Klimcik, P. Presnajder, Int. J. Theor. Phys. 35, 231
(1996)

\bibitem{ElizaldePistons} E. Elizalde, S. D. Odintsov, A. A.
Saharian \arx0902.0717

\bibitem{KirstenPistons} S. A. Fulling, K. Kirsten, \plb671, 179
(2009)

\bibitem{nam} S. Nam, JHEP, 10, 044 (2000)

\bibitem{malakas} A. Edery, Phys. Rev. D75, 105012 (2007)

\bibitem{marakefsky} V. Marachevsky, Phys. Rev. D75, 085019 (2007)

\bibitem{hybrid} X.-H. Zhai, X.-Z. Li, Phys. Rev. D76, 047704 (2007)

\bibitem{calvacanti} R. M. Cavalcanti, \arx quant-ph/0310184

\bibitem{fulling} S. A. Fulling, J. H. Wilson, quant-ph/0608122

\bibitem{vasilis} V. K. Oikonomou, arXiv: 0903.4159

\bibitem{perivolaropoulos} L. Perivolaropoulos, Phys. Rev. D77,
107301 (2008)

\bibitem{kinezoi} Xiang-Huazhai, Yan-Yanzhang, Xin-Zhouli,
arXiv:0808.0062

\bibitem{Bordagreview} M. Bordag, U. Mohideen, V. M. Mostepanenko,
Phys. Rep. 353, 1 (2001)

\bibitem{elizalde} E. Elizalde, "Ten physical
applications of spectral zeta functions", Springer (1995); E.
Elizalde, S.~D.~Odintsov, A. Romeo, A. A. Bytsenko, "Zeta
regularization techniques and applications", World Scientific
(1994)

\bibitem{kirstenbook} Klaus Kirsten, Spectral Functions in Mathematics and
Physics,  Chapman Hall/CRC (2001)

\bibitem{kirsten12}K. Kirsten, Generalized multidimensional Epstein zeta functions, J. Math. Phys. 35, 459-470 (1994)

\bibitem{kirsten14}K. Kirsten,  Inhomogeneous multidimensional Epstein zeta functions, J. Math. Phys. 32, 3008-3014 (1991)

\bibitem{Teo} L. P. Teo, \arx0812.4641; L. P. Teo, \arx0901.2195

\bibitem{TeoMassive} S. C. Lim, L. P. Teo, \arx0807.3631

\bibitem{chodos} T. Appelquist, A. Chodos, Phys. Rev. Lett. 50 141
(1983)

\bibitem{gradsteyn} I.S. Gradshteyn and I.M. Ryzhik, Table of Integrals Series and Products
(Academic Press, 1965)



\end{thebibliography}
\end{document}